\documentclass[12pt,twocolumn]{article}
\usepackage[a4paper,top=0.6in,left=0.6in, footskip=0.5in,marginparwidth=0.1in]{geometry}

\usepackage{float}

\usepackage[utf8]{inputenc}
\usepackage{cite}

\usepackage{lastpage,fancyhdr,graphicx}
\usepackage{nameref}
\usepackage{hyperref}
\RequirePackage{jabbrv}
\usepackage{microtype}
\DisableLigatures[f]{encoding = *, family = * }
\usepackage{changepage}
\usepackage[aboveskip=15pt,font=small,labelfont=bf,labelsep=period,singlelinecheck=off]{caption}
\makeatletter
\renewcommand{\@biblabel}[1]{\quad#1.}
\makeatother

\usepackage{multicol}
\usepackage{epstopdf}

\usepackage{siunitx}
\DeclareSIUnit{\Molar}{M}
\DeclareSIUnit{\Osmole}{Osm}
\usepackage{amssymb}
\usepackage{multirow}

\usepackage{bm}
\usepackage{booktabs,caption}
\usepackage[flushleft]{threeparttable}
\usepackage{authblk}

\usepackage{color}
\definecolor{Gray}{gray}{.25}
\usepackage{graphicx}

\usepackage{wrapfig}
\usepackage[pscoord]{eso-pic}
\usepackage[fulladjust]{marginnote}
\reversemarginpar

\author[a]{Lucas Le Nagard}
\author[a]{Aidan T. Brown} 
\author[a]{Angela Dawson} 
\author[a]{Vincent A. Martinez} 
\author[a,1]{Wilson C. K. Poon} 
\author[b,1]{Margarita Staykova}
\affil[a]{\small{School of Physics and Astronomy, The University of Edinburgh, King's Buildings, Peter Guthrie Tait Road, Edinburgh EH9 3FD, United Kingdom.}}
\affil[b]{\small{Department of Physics, Durham University, South Road, Durham DH1 3LE, United Kingdom.}}
\affil[1]{\small{E-mail: w.poon@ed.ac.uk, margarita.staykova@durham.ac.uk}}

\title{{\Large
\textbf{Encapsulated bacteria deform lipid vesicles into flagellated swimmers}
}}
\date{\vspace{-5ex}}

\begin{document}

\twocolumn[
  \begin{@twocolumnfalse}
\maketitle

\noindent{\bf \Large Significance:} Swimming bacterial pathogens can penetrate and shape the membranes of their host cells. We study an artificial model system of this kind comprising \textit{Escherichia coli} enclosed inside vesicles, which consist of nothing more than a spherical membrane bag. The bacteria push out membrane tubes, and the tubes propel the vesicles. This phenomenon is intriguing because motion cannot be generated by pushing the vesicles from within. We explain the motility of our artificial cell by a shape coupling between the flagella of each bacterium and the enclosing membrane tube. This constitutes a design principle for conferring motility to cell-sized vesicles, and demonstrates the universality of lipid membranes as a building block in the development of new biohybrid systems.

\vspace*{0.1in}

\end{@twocolumnfalse}
]

\medskip

{\bf We study a synthetic system of motile \textit{Escherichia coli} bacteria encapsulated inside giant lipid vesicles. Forces exerted by the bacteria on the inner side of the membrane are sufficient to extrude membrane tubes filled with one or several bacteria. We show that a physical coupling between the membrane tube and the flagella of the enclosed cells transforms the tube into an effective helical flagellum propelling the vesicle. We develop a simple theoretical model to estimate the propulsive force from the speed of the vesicles, and demonstrate the good efficiency of this coupling mechanism. Together, these results point to design principles for conferring motility to synthetic cells.}

\medskip

The interaction of active and passive matter lies at the heart of biology. Active matter~\cite{Ramaswamy2010} consists of collections of entities that consume energy from their environment to generate mechanical forces, which often result in motion. Thus the cell membrane, whose essential component is a passive lipid bilayer, can actively remodel during cell growth, motility and, ultimately, cell evolution, under the forces exerted by a host of active agents. For example,  continuous polymerization-depolymerization of actin in the cytoskeleton deforms the eukaryotic cell membrane into two- and one-dimensional protrusions (lamellipodia and filopodia) that can move the whole cell~\cite{blanchoin2014actin,koster2016cortical}. Similar actin-supported membrane protrusions are thought to have facilitated the accidental engulfment of bacteria that led to the emergence of eukaryotic cells~\cite{yutin2009}. The biophysics of active membranes has therefore been subjected to interdisciplinary scrutiny~\cite{Mouritsen2016}. More recently, learning how to create active membranes systems that deform, divide and propel has become a priority area in the drive to synthesize life \textit{ab initio}~\cite{Schwille2018}.

Lipid vesicles enclosing natural or artificial microswimmers are becoming a model system for studying active membranes \textit{in vitro} ~\cite{paoluzzi2016shape,chen2017rotational,wang2019shape,shan2019assembly,quillen2020boids,li2019shape,peng2022activity,takatori2020active,vutukuri2020active}. Such composites have also direct biological relevance. For instance, from inside their eukaryotic hosts, bacterial pathogens such as \textit{Rickettsia rickettsii} or \textit{Listeria monocytogenes}~\cite{Ortega2019,dowd2020listeria} continue their life cycles by hijacking the actin polymerization-depolymerization apparatus of their hosts and pushing out a tube-like protuberance from the plasma membrane.  The pathogens then contact other host cells or escape into the surrounding medium by means of these membrane tubes~\cite{Colonne2016}. 

To date, research in coupling swimmers with membranes has mostly been theoretical and numerical. Such models have predicted a range of interfacial morphological changes, and in some cases, net motion of the interface~\cite{paoluzzi2016shape,chen2017rotational,wang2019shape,shan2019assembly,quillen2020boids,li2019shape,peng2022activity}. The experimental realization of these systems was only recently achieved by encapsulating swimming \textit{Bacillus subtilis} bacteria~\cite{takatori2020active} and synthetic Janus particles~\cite{vutukuri2020active} in giant lipid vesicles. Both experiments reported non-equilibrium membrane fluctuations and vesicle deformations, ranging from tubular protrusions to dendritic shapes. However, net motion of the vesicles was not observed in either case. 

Here, we present a similar experimental design but with markedly different outcome. \textit{Escherichia coli}, another common motile bacterium, also extrudes membrane tubes but in addition sets the whole vesicle into motion. We demonstrate that such motion is due to a physical coupling between the flagella bundle of the enclosed cells and the tubes. The tube-flagella composite functions as a helical propeller for the entire vesicle. 

In biology, the specificity of interactions between bacteria and the membranes of eukaryotic hosts underlies the plethora of parasitic and symbiotic relations that have emerged between cells~\cite{Colonne2016,martinez2018tiny,liss2015take}. Likewise, our observations illustrate the importance of small details in the design of active matter systems~\cite{Brown2016}. Encapsulated bacteria propelled by a single bundle of helical flagella can generate net motion of the vesicles, whereas encapsulated swimmers propelled at similar speeds by phoresis fail to do so~\cite{vutukuri2020active}. These observations illustrate the fact that it is dangerous to proceed from coarse-grained simulations or theory that neglect such details to predict the behavior of particular systems. At the same time, our results point to a design principle for conferring motility to artificial cell models.

\begin{figure*}[t]
\centering
\includegraphics[width= 0.8\textwidth]{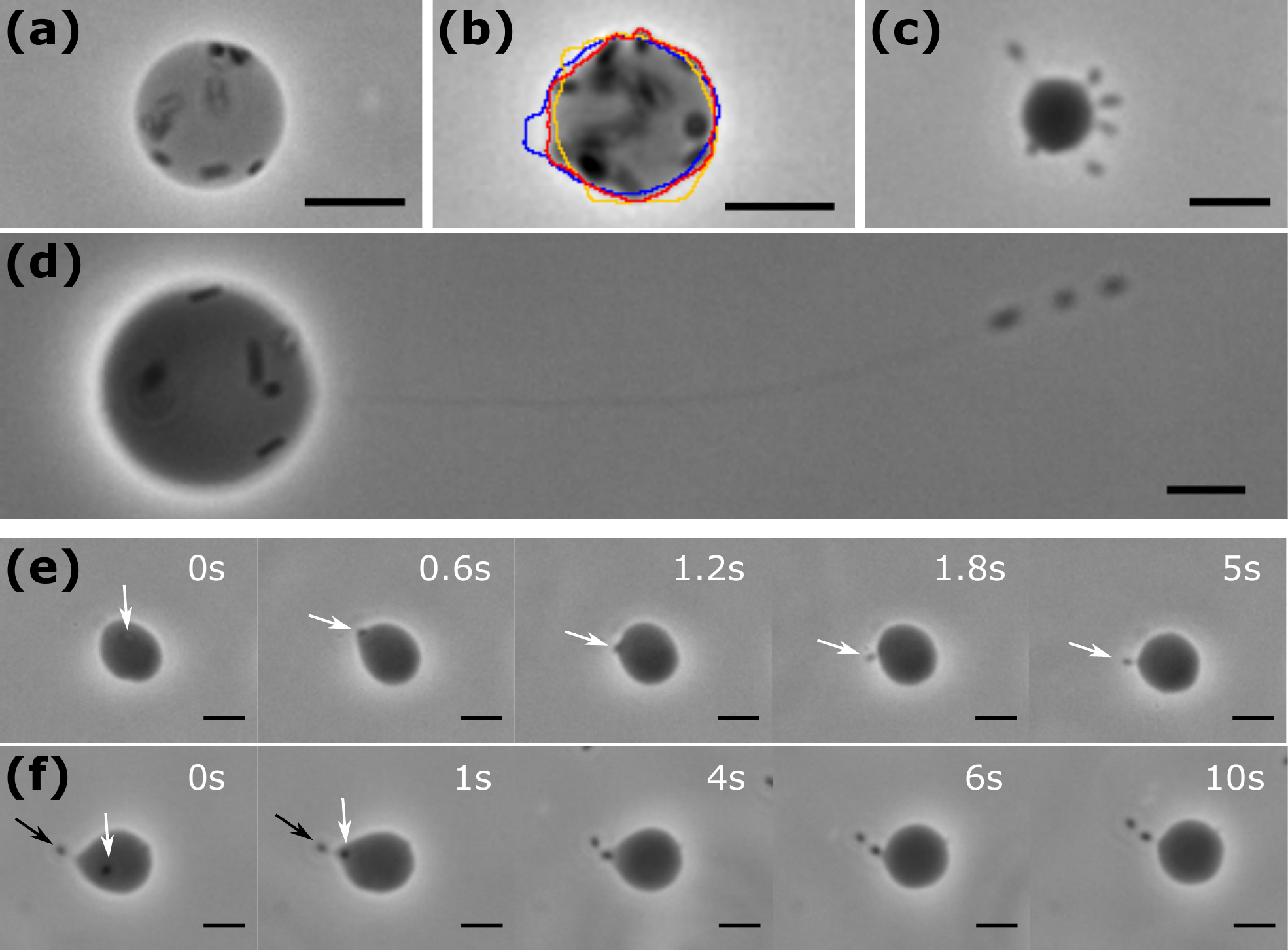}
\caption{\label{Fig1}\textbf{Bright-field images of GUVs in the course of osmotic deflation displaying membrane fluctuations and tubular protrusions.} \textbf{(a)} Tense GUV encapsulating approximately 10 bacteria (dark and lighter spots within the GUV) and displaying no visible fluctuations. \textbf{(b)} Deflated GUV displaying bacteria-amplified fluctuations illustrated by three superimposed contours. \textbf{(c)} GUV displaying 6 tubular protrusions extruded by bacteria, each containing a single cell. \textbf{(d)} GUV displaying a long ($\gtrsim$\SI{100}{\micro\metre}) tube with three cells at its extremity. \textbf{(e)} Image sequence showing a single cell (white arrow) extruding a tube by pushing on the membrane. \textbf{(f)} Image sequence recorded on a different GUV, showing a second cell (white arrow) entering a tube previously extruded by a single bacterium (black arrow), which leads to elongation of the tube. (Scale bars, 10~\si{\micro\metre}.)}
\end{figure*}

\section*{Results}

\subsection*{Encapsulated bacteria extrude lipid tubes}

We encapsulated a smooth-swimming strain of K-12 derived \textit{E.~coli} in giant unilamellar vesicles (GUVs) made of 1-palmitoyl-2-oleoyl-sn-glycero-3-phosphocholine lipids (POPC) and doped with a fluorescent dye using the inverted-emulsion method~\cite{pautot2003production}. The internal medium was lysogeny broth (LB) supplemented with sucrose, which the bacteria cannot metabolise. The external medium was an aqueous solution of glucose, which, like sucrose, does not diffuse across the lipid membrane. Our GUVs sedimented towards the glass bottom of the sample chambers, which had been pre-treated with bovine serum albumin to minimise vesicle adhesion. Bacteria and vesicles were imaged with an inverted microscope in phase-contrast and fluorescence modes. We occasionally observed cell division in the GUVs and the bacteria remained motile for at least $\sim\SI{8}{\hour}$, thanks to nutrients provided in the inner medium and/or endogenous metabolism~\cite{schwarz2016escherichia} enabled by the diffusion of dissolved O$_2$ through the membrane. 

In sample chambers sealed immediately after bacteria encapsulation, the majority of GUVs appeared tense and spherical, with no visible sign of shape fluctuations, Fig.~\ref{Fig1}(a). The $\sim 10$ encapsulated bacteria typically swam just below the inner surface, reminiscent of previous observations of \textit{E.~coli} in lecithin-stabilised spherical water-in-oil emulsion droplets~\cite{vladescu2014filling}. Amongst these GUVs, we occasionally observed GUVs with tubular protrusions containing one or more bacteria. 

To increase the number of GUVs exhibiting such protrusions, we used osmotic deflation to decrease the membrane tension of our vesicles by keeping sample chambers open for a defined period of time, $t_{\rm w}$, before sealing for observation. Water evaporation increased the osmolarity of the external solution by a controlled factor $\alpha$ (see Methods), leading to water efflux. Progressive deflation yielded a higher fraction of GUVs with bacteria-containing tubes than sudden mixing with a hypertonic buffer. At $t_{\rm w} = 0$, the internal and external media were isotonic ($\alpha=1$), giving mostly tense vesicles with only a fraction $\chi = 0.17$ showing bacteria in tubular protrusions (see SI for details). At even a moderate deflation, $\alpha = 1.05$, some vesicles exhibited pronounced fluctuations at $t_{\rm w} \approx \SI{20}{\min}$ and $\chi = 0.49$. Further deflation, $\alpha = 1.1$, gave $\chi = 0.67$, $t_{\rm w} \approx \SI{40}{\min}$, with tube-bearing vesicles dominating our field of view, Fig.~\ref{Fig1}(c). The vesicles displayed either multiple tubes, Fig.~\ref{Fig1}(c), or a single tube with one or multiple bacteria inside, Fig.~\ref{Fig1}(d). 

\begin{figure*}[t!]
\centering
\includegraphics[width=0.8\textwidth]{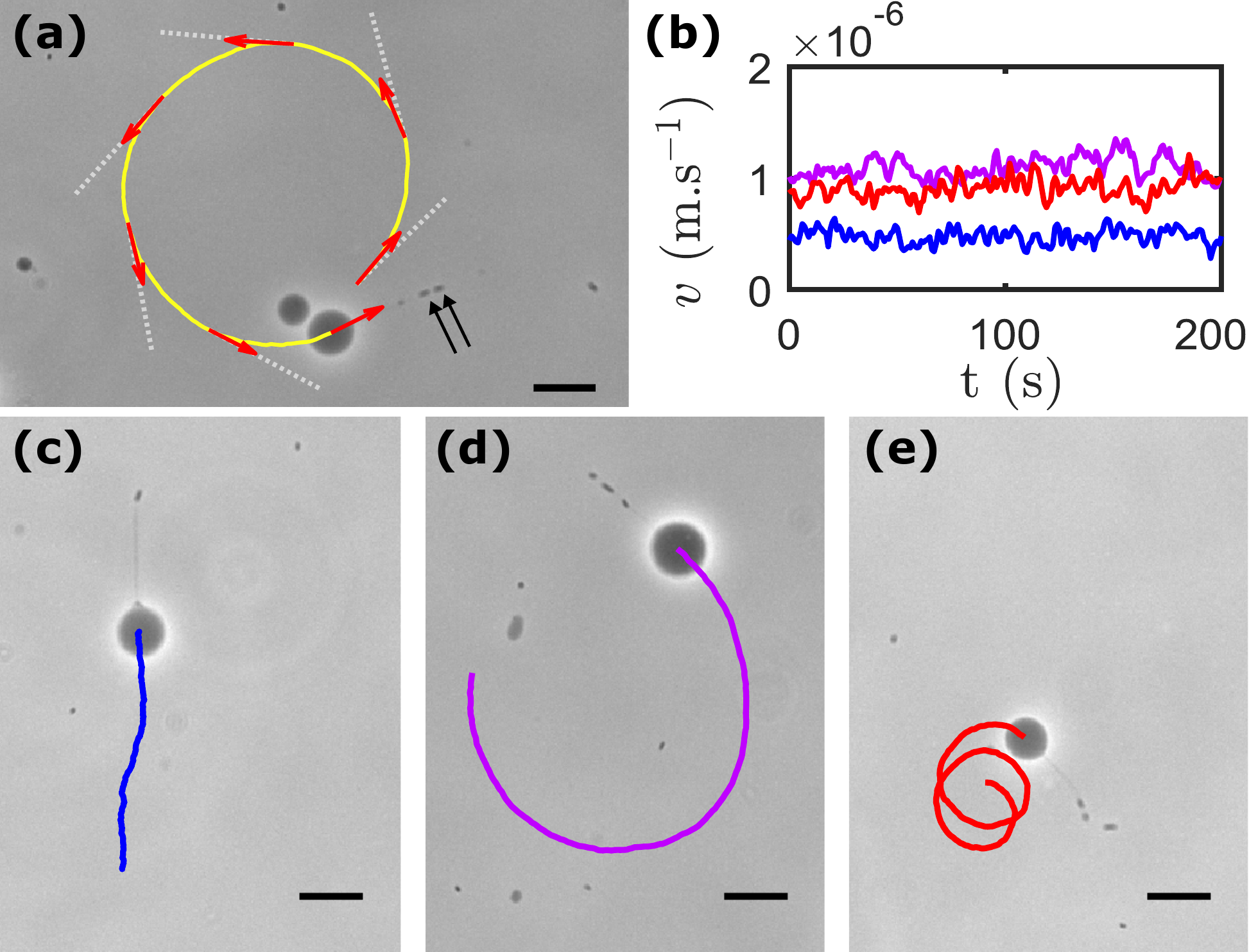}
\caption{\label{Fig2}\textbf{Propulsion of GUVs containing \textit{E. coli} cells.} \textbf{(a)} Trajectory (yellow line, $\Delta t=\SI{180}{\second}$) of a GUV propelled by two bacteria in a membrane tube (black arrows). Red arrows indicate the orientation of the instantaneous velocity vector at \SI{30}{\second} intervals, and white dotted lines indicate the orientation of the tube at the same time points, showing that the GUV swims tube-first. \textbf{(b)} Speed as a function of time for the three GUVs displayed in (c)-(e), with matching colors. \textbf{(c-e)} Vesicle trajectories recorded over \SI{200}{\second} may vary from \textbf{(c)} straight line, \textbf{(d)} counterclockwise to \textbf{(e)} clockwise rotation. (Scale bars, 20~\si{\micro\metre}.)
}
\end{figure*}

We were able to capture occasionally the rapid process of the initiation of tubular protuberances by swimming bacteria. This takes approximately \SI{1}{\second} and requires a cell to swim perpendicular to the membrane, Fig.~\ref{Fig1}(e). A second cell can enter an already formed tube, Fig.~\ref{Fig1}(f), causing its extension. Tube elongation consumes membrane and generates tension, Fig.~\ref{Fig1}(f), resulting in a more spherical shape and preventing further tube growth. Interestingly, a bacterium swimming into a pre-existing cell-free membrane tube was never observed. Moreover, no bacteria-containing tubes were seen with encapsulated dead cells, and bacteria swimming outside vesicles do not pull off membrane tubes under our conditions (see SI).

\subsection*{Bacteria in membrane tubes propel lipid vesicles}

In striking contrast to GUVs encapsulating Janus particles or \textit{Bacillus subtilis} bacteria, which display large deformations but remain static~\cite{vutukuri2020active,takatori2020active}, we observed that motile \textit{E. coli} in membrane tubes were able to propel GUVs at typical speeds $v\sim\SI{1}{\micro\metre\per\second}$, Fig.~\ref{Fig2} and Movie~S1. The motion is always tube-first with velocity vector parallel to the tube, Fig.~\ref{Fig2}(a,~c-e). GUVs with bacteria but no tubular protrusions remain static, suggesting that motile bacteria in the vesicle lumen do not contribute to vesicle propulsion.

We found that vesicle trajectories varied from straight to curved clockwise (CW) and counterclockwise (CCW) within the same sample (viewed from the fluid side), Fig.~\ref{Fig2}(c)-(e). Out of all assessed GUV trajectories (65 in total), $34\%$ were curved CW, $42\%$ were curved CCW, $20\%$ were straight and $5\%$ displayed a reversal of their curvature during the tracking (the latter usually triggered by some rearrangement of the bacteria in the tube). Vesicle trajectories therefore tend to be curved, with no strong preference between CW and CCW rotation. This contrasts with unencapsulated bacteria outside vesicles, for which only $7\%$ of circular trajectories near the glass slide (10 out of 146) were CCW. Such behaviour, governed by hydrodynamic interactions between the bacteria and the glass substrate, agrees with previous measurements~\cite{lemelle2013curvature} and is consistent with a no-slip boundary condition operating at the slide~\cite{lauga2006swimming}. It is clear that these strong hydrodynamic effects do not translate into biased rotation of our vesicle-bacteria system, which is unsurprising given the more complex and variable geometry in that case.

\subsection*{A physical coupling between flagella and membrane tubes generates a propulsive force}

\begin{figure*}[t]
\centering
\includegraphics[width= 0.7\textwidth]{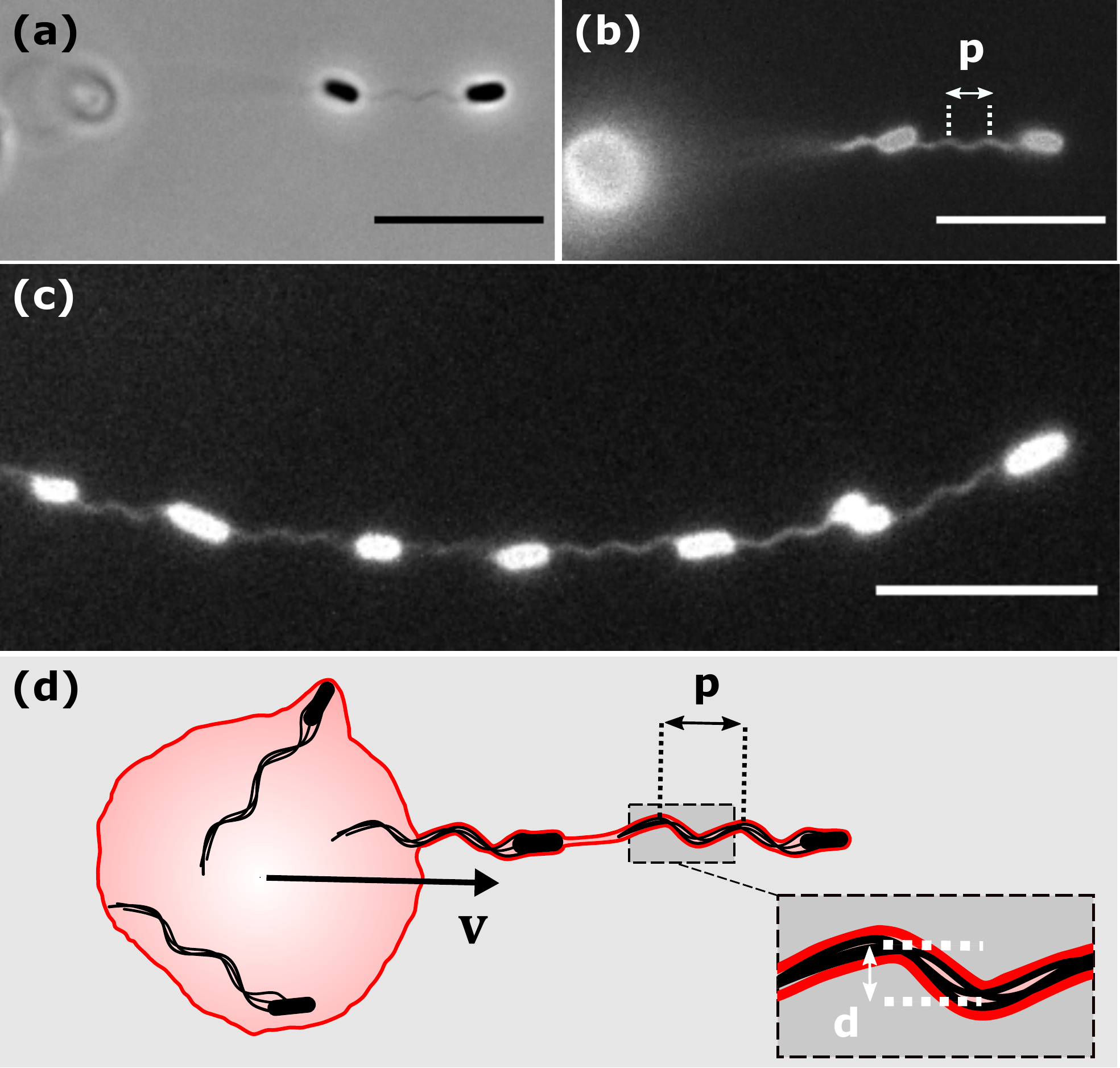}
\caption{\label{Fig3}\textbf{Membrane tubes propel GUVs by tightly coupling with the flagella bundles of enclosed cells.} \textbf{(a)} Phase contrast image of two cells in a tube. The out-of-focus GUV is barely visible on the left. Close inspection reveals that the portion of tube connecting the two cells is helical. \textbf{(b)} Fluorescence image of the lipid tube displayed in (a), confirming its helical shape with pitch $p=\SI{2.3}{\micro\metre}$. The fluorescence signal is emitted by the dye embedded in the GUV membrane. The bright disk on the left is the out-of-focus GUV. The tube appears blurred behind the second cell in the tube (\textit{i.e.} closer to the GUV) in (a) and (b) due to the small depth of focus. \textbf{(c)} Fluorescence image of a tube containing multiple bacteria, showing the coupling with the flagella bundles behind each cell. \textbf{(d)} Schematic of the system (not to scale) describing the mechanism of GUV propulsion. The membrane (red contour) of the tube wraps the bacteria and adopts the shape of their helical flagella (pitch $p$, helical diameter $d$). Flagellar rotation within the tube generates a thrust force and results in an instantaneous velocity vector $\mathbf{v}$ parallel to the tube. (Scale bars, 10~\si{\micro\metre}.)
}
\end{figure*}

The self-propelled motion of our biohybrid vesicles is a surprising phenomenon, because encapsulated bacteria are isolated from the external medium by a lipid membrane. To get insights into its mechanisms, we turn to high resolution microscopy. Imaging the membrane directly by combining phase contrast microscopy with fluorescent imaging, Fig.~\ref{Fig3}, shows bacteria  tightly wrapped in tubes with $R_{\rm t}\lesssim \SI{0.5}{\micro\metre}$ along the whole length of any tube. The portion of tube behind each bacterial body is also noticeably thinner than the cell body itself. Knowing that the nearly rigid helical flagella bundle in swimming \textit{E.~coli} has a helical diameter $d$ comparable to that of the cell body~\cite{turner2000real}, this observation suggests that the portion of tube surrounding the flagella bundle is severely distorted from a cylindrical form.

Our images indeed show that the two dimensional projection of the membrane around the flagella bundle has a sinusoidal shape, Fig.~\ref{Fig3}(a)-(c). We fitted the images of 10 helical tubes to a sine function, Fig.~S2, and determined the pitch $p=2.3\pm\SI{0.2}{\micro\metre}$ and helical diameter $d=0.4\pm\SI{0.1}{\micro\metre}$ (mean $\pm$ standard deviation), which agree with previous values for the flagella bundle of \textit{E. coli}~\cite{turner2000real}. High-speed imaging returned rotation frequencies of typically 50 to \SI{120}{\hertz} (e.g., \SI{90}{\hertz} for the tube in Fig.~\ref{Fig3}(b)), which again falls within the range measured for the flagella bundle of \textit{E.~coli}~\cite{chattopadhyay2006swimming} (Movie~S2). We conclude that the lipid membrane wraps closely enough around the flagella bundle to become helical, and that the bundle retains the same geometry as in free-swimming cells.

These findings suggest a propulsion mechanism that is most easily applied to the simplest, single-tube vesicle, Fig.~\ref{Fig3}(d). The thin membrane tube couples to the rotating helical flagella bundle of the nested bacterium, adopts its shape, and undergoes helical motion, which generates a thrust force. Consistent with this, we expect the propulsive force to be proportional to the number of bacteria in the tube if cell-cell interactions remain negligible.

\subsection*{The propulsive force scales with the number of bacteria in a tube}

\begin{figure*}[t]
\centering
\includegraphics[width= 0.68\textwidth]{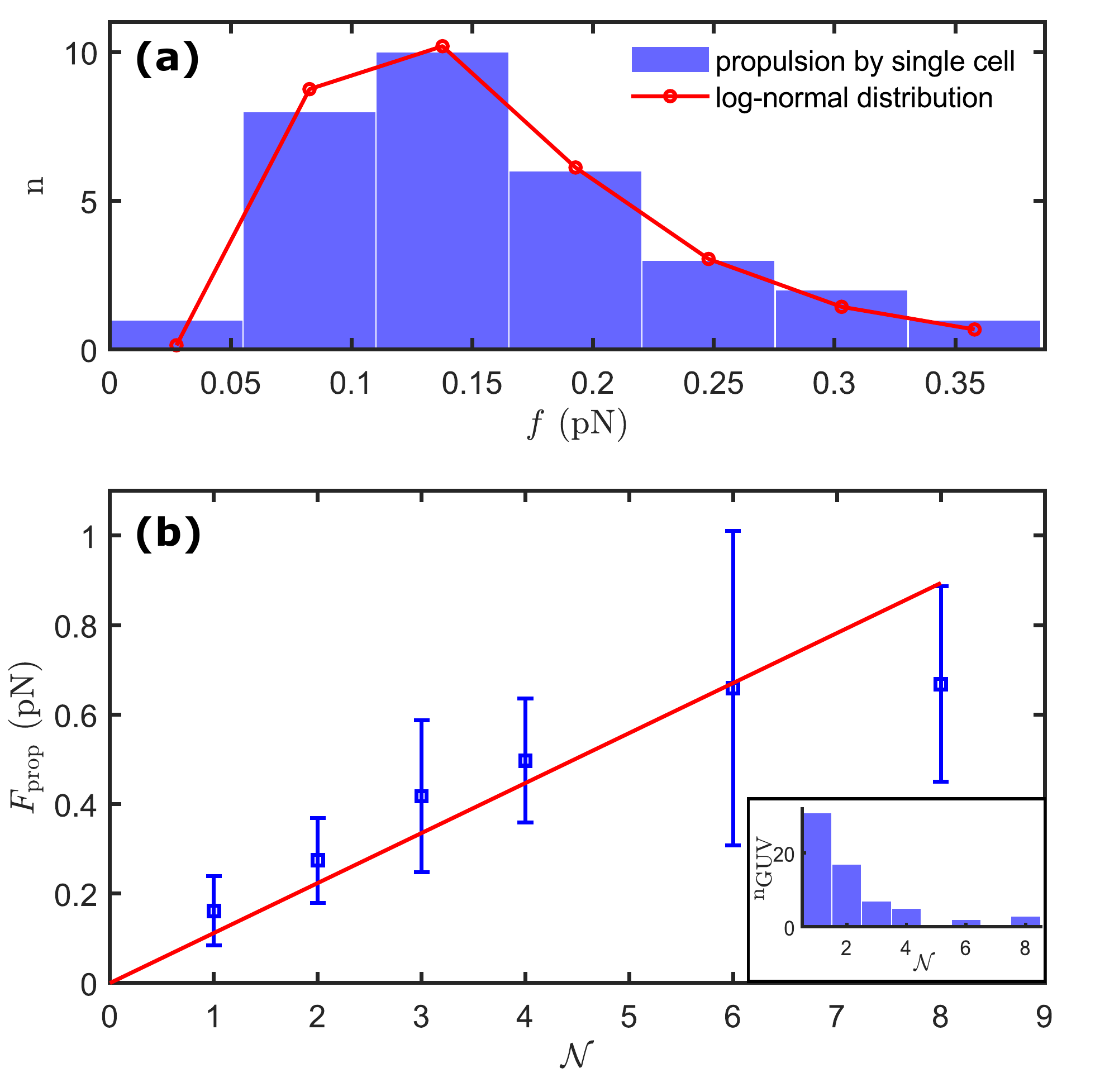}
\caption{\label{Fig4} \textbf{Propulsive force generated by bacteria in membrane tubes.} \textbf{(a)} Histogram of $f$, the propulsive force estimated for all GUVs propelled by a single cell in a tube ($n=31$). The distribution is described by a log-normal probability density function: ${\rm P}(f)=\frac{1}{\sqrt{2\pi}f\sigma}{\rm exp}\left(-\frac{({\rm ln}(f)-\mu)^2}{2\sigma^2}\right)$ with parameters $\mu=-1.9$, and $\sigma=0.48$ for $f$ in $\si{\pico\newton}$ (converted to discrete values, red line), leading to a mean value $\overline{f} =\num{0.16}~\si{\pico\newton}$. \textbf{(b)} Average value of the total propulsive force $F_{\rm prop}$ generated by bacteria as a function of the number of bacteria $\mathcal N$ in the tubes. The red line is a linear fit weighted by the inverse of the variance of the data, with a fitted slope $\langle f \rangle =\num{0.11}\mathcal{N}~\si{\pico\newton}$ ($R^2=0.8$). Error bars correspond to standard deviations. Inset: number $n_{\rm GUV}$ of GUVs tracked for each value of $\mathcal N$.
}
\end{figure*}

To verify this, we consider motile vesicles propelled by a single bacteria-bearing membrane tube and exclude cases such as that in Fig.~\ref{Fig1}(c). The relevant Reynolds number is Re~$=2Rv\rho_{\footnotesize{\rm os}} /\eta\sim10^{-5}$, where $R\sim\SI{10}{\micro\metre}$ is the GUV radius, $\rho_{\footnotesize{\rm os}}$ is the volume mass of the outer solution and $\eta\sim\SI{1.5}{\milli\pascal\second}$ is its viscosity at \SI{20}{\celsius}~\cite{chirife1997simple,telis2007viscosity}. Inertia is therefore negligible, and the propulsive force generated by the bacteria in a membrane tube, $F_{\rm prop}$, is exactly balanced by the total drag force, $F_{\rm drag}=\xi v$, acting on the composite swimmer (GUV + tube), where $\xi$ is its friction coefficient and $v$ its speed:
\begin{equation}
\label{Eq1}
  F_{\rm drag}=\xi v=F_{\rm prop}
\end{equation}

We include the contribution of the GUV and of the membrane-wrapped bacteria to $\xi$, but neglect the much thinner, `empty' portions of tube (see SI and Fig.~S5). We also neglect wall drag because we cannot quantify the GUV-wall distance. Thus, for a GUV bearing a single $\mathcal N$-bacteria tube, $\xi=\xi_v+\mathcal{N}\xi_b$, with $\xi_v$ and $\xi_b$ the friction coefficients of the spherical GUV and of a single membrane-wrapped \textit{E.~coli} respectively. A non-slip boundary condition applies to GUVs in flow~\cite{jahl2020lipid}, so that for a spherical GUV of radius $R$,  $\xi_v=6\pi\eta R$. Modelling \textit{E. coli} cells as oblate ellipsoids of minor and major axes $a$ and $b$ leads to $\xi_b=\frac{2\pi\eta b}{\ln{\left(2b/a\right)}-1/2}$~\cite{schwarz2016escherichia}. We therefore have 
\begin{equation}
\label{Eq2}
  \xi=6\pi\eta R+\frac{2\mathcal{N}\pi\eta b}{\ln{\left(2b/a\right)}-1/2}.
\end{equation}
We take $a=\SI{1}{\micro\metre}$ and $b=\SI{3}{\micro\metre}$ throughout.

\begin{figure*}[t]
\centering
\includegraphics[width= 0.75\textwidth]{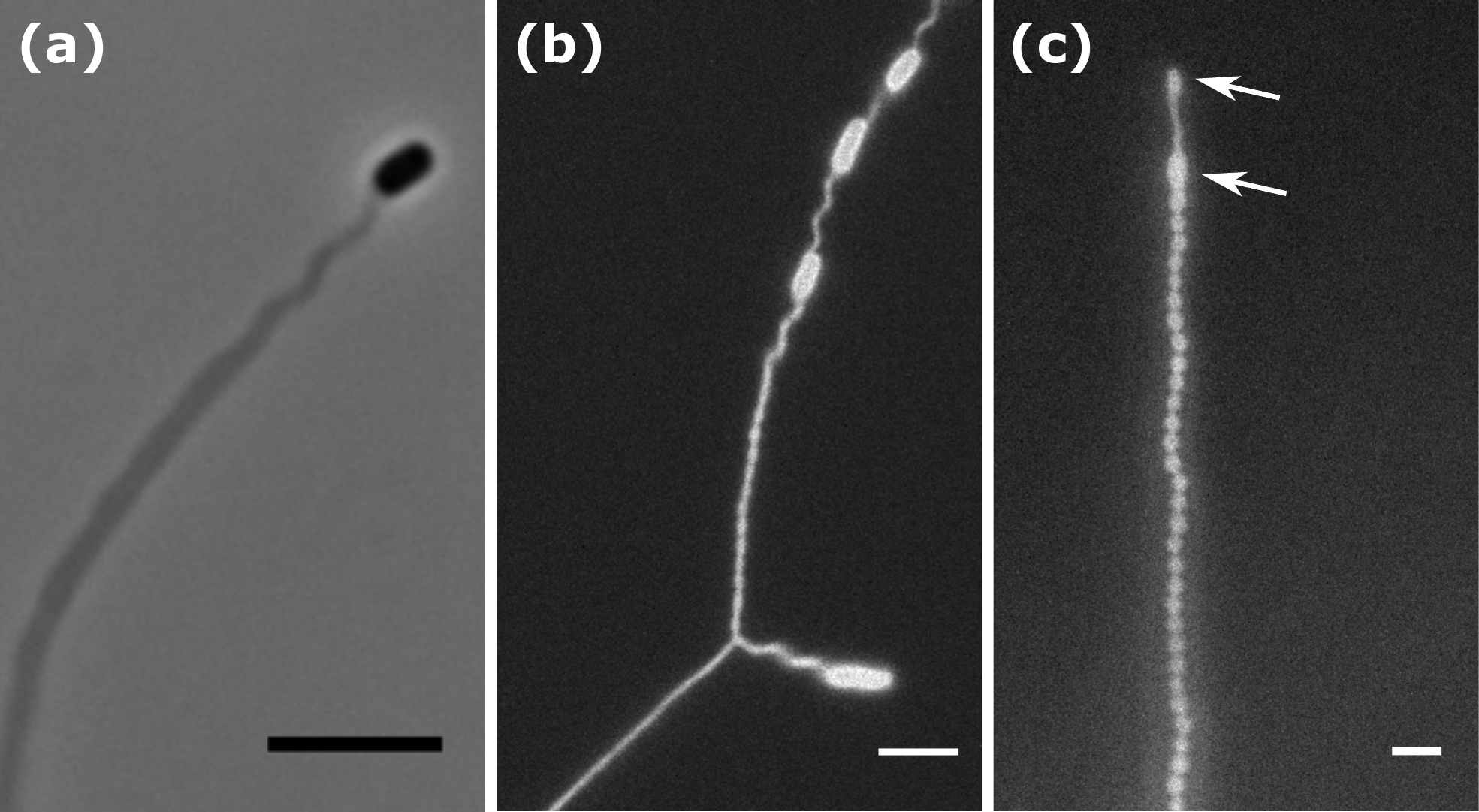}
\caption{\label{Fig5} \textbf{Thicker, split and pearled tubes.} \textbf{(a)} Thicker tube coupling with the flagella bundle of an enclosed cell. \textbf{(b)} Tube split into two sub-tubes, one containing a single cell and the other one containing 4 bacteria (only 3 are visible). \textbf{(c)} Pearled tube with two bacteria at its end (white arrows). (Scale bars, \SI{5}{\micro\metre}.)
}
\end{figure*}

Consider first only vesicles propelled by a single cell in a tube, for which we define $F_{\rm prop}(\mathcal N=1)=f$. Applying Eq.~\ref{Eq1} using measured vesicle speeds and calculated $\xi$ as inputs, we find a range of $f$ values, Fig.~\ref{Fig4}(a). Unlike $v$, we expect the propulsive force to be independent of the vesicle radius, and indeed we find no correlation between $f$ and $R$, Fig.~S3. Instead, the observed variability in $f$ is likely due to the distribution of single-bacterium swimming speeds (as revealed by differential dynamic microscopy, Fig.~S4 and ref.~\cite{wilson2011differential}) and variable (but unknown) vesicle-wall hydrodynamic interactions. Variations in tube radius around the flagella bundle are predicted to contribute only weakly to this variability (see SI and Fig.~S6). The spread in $f$ is well-described by a log-normal distribution with a mean of \SI{0.16}{\pico\newton}, about three times lower than the thrust generated by a free-swimming \textit{E.~coli}~\cite{chattopadhyay2006swimming,drescher2011fluid}. However, since we neglect wall drag~\cite{malysa1986rotational}, our estimate of $f$ is a lower bound. The true mean $f$ may therefore be statistically indistinguishable from the propulsive force of a free-swimming cell, revealing a surprisingly efficient coupling between the encapsulated flagella with the external fluid via the surrounding lipid membrane.

Turning to multi-bacteria tubes, we used Eq.~\ref{Eq1}~\&~\ref{Eq2}, and deduced $F_{\rm prop}$ for GUVs propelled by tubes containing up to eight bacteria,  Fig.~\ref{Fig4}(b). There is significant variability in the deduced values, especially at large $\mathcal N$. The sources of variability previously mentioned for the distribution in single-cell $f$ values continue to be relevant here, with potentially the addition of variable cell-cell interactions. Moreover, only a small number of GUVs were tracked for $\mathcal N\geq 6$ ($n_{\rm GUV}\leq 3$ in each case), because larger GUVs with longer tubes lead to a higher probability of them interacting with other GUVs, precluding their use for data collection. 

With these caveats, we find $F_{\rm prop} =\mathcal{N} \langle f \rangle$, with a fitted  $\langle f \rangle  = \SI{0.11}{\pico\newton}$ (regression coefficient~= 0.8 with the origin included as a data point), consistent with the single-cell force estimate of \SI{0.16}{\pico\newton}. Such proportionality confirms that individual bacteria in a membrane tube are independent sources of propulsion, with negligible  hydrodynamic and other interaction between them.

\section*{Discussion}

The tube size appears to be an important factor for the propulsion of vesicles. GUVs with very short tubes will not move because flagella bundles cannot contribute to the propulsion from the vesicle lumen, Fig.~\ref{Fig3}(d). On the other hand, only tubes that are sufficiently thin will couple to the shape of the rotating flagella and undergo a helical motion. Indeed, the majority of our observations depict tubes closely wrapped around the flagella bundles, but in some rare cases we also observed thicker thrust-generating tubes (diameter $\sim\SI{1}{\micro\metre}$), Fig.~\ref{Fig5}(a) and Movie~S3. This implies that a tight, no-slip coupling of flagella bundle and membrane is not essential to the propulsion; flagella can generate a propulsive deformation of the membrane also through an interstitial aqueous layer. Our considerations are supported by theoretical studies predicting same propulsion speeds for flagella undergoing a rigid helical rotation and a helical wave deformation ~\cite{Powers14}. In either case, as long as the membrane tube is thin enough to be perturbed by the enclosed rotating bundle, it adopts its helical shape and functions as an `effective flagellum' generating propulsion for the whole vesicle. 

Interestingly however, the thin tubes in our experiments do not agree with the classical elastic theory of membrane tubulation ~\cite{derenyi2002formation}, which predicts tube radii $R_t=\frac{2\pi\kappa}{F}\sim\SI{1.3}{\micro\metre}$, where $\kappa\sim 10^{-19}~\si{\joule}$ is the bending rigidity of a POPC membrane~\cite{dimova2014recent}, and $F\approx \SI{0.5}{\pico\newton}$ is the thrust force generated by a free-swimming \textit{E. coli}~\cite{chattopadhyay2006swimming,drescher2011fluid}. 

Such discrepancies are usually reconciled by the existence of a positive spontaneous membrane curvature~\cite{lipowsky2013spontaneous} and/or a relative excess of lipids in the outer membrane leaflet~\cite{bozic1992role,miao1994budding}. Both favor thin membrane protrusions~\cite{glassinger2006influence} and are consistent with our observations of spontaneously occurring thin tubes after deflation, even in bacteria-free GUVs. Control experiments suggest that a transmembrane lipid density asymmetry is a more likely mechanism (see SI and Fig.~S7). The inverted emulsion encapsulation protocol offers a much longer time for outer leaflet equilibration than for the inner leaflet which, combined with the slow adsorption of POPC lipids to oil/water interfaces ($\gtrsim\SI{1}{h}$ to reach saturation)~\cite{pautot2003production}, leads us to expect a difference in the lipid densities between the two leaflets. Vutukuri et al. used a similar method to produce GUVs and observed that their self-propelled Janus particles were able to extrude membrane tubes at much lower propulsive forces than the theoretically predicted tubulation force~\cite{vutukuri2020active}. Though likely to be of general interest, a detailed investigation of this issue and the physics associated with GUV production by the inverted emulsion method are beyond our scope.

For completeness, we also report here some rarer tube morphologies, which illustrate the richness of behaviours spontaneously occurring in our system. Bacteria can sometimes generate split tubes, Fig.~\ref{Fig5}(b), with dynamic junctions moving along the tube, Movie~S4. Pearling instabilities can also be observed, especially in thicker tubes, behind the section wrapping the flagella bundles, Fig.~\ref{Fig5}(b, c). Occasionally, cells divide in the tube, sometimes leading to interesting tube dynamics, Movie~S5.

\section*{Conclusion}

We showed that motile \textit{E.~coli} swimming inside deflated GUVs can deform the lipid membrane and extrude tubular membrane protrusions, in agreement with recent studies using \textit{B.~subtilis} and synthetic Janus particles~\cite{vutukuri2020active, takatori2020active}. Whereas previous systems remained static, our vesicles display self-propelled motility. Detailed microscopy revealed that the membrane tube wraps the enclosed bacteria and follows the motion of their helical flagella bundle. Such a tube acts as an effective flagellum for the vesicle and provides the propulsive force needed to generate a sustained motion. In passing, we note that some bacteria, such as various Gram-negative \textit{Vibrio} species, have evolved to produce a membranous sheath around their flagella, the role of which remains poorly understood~\cite{zhu2017molecular}. Future biophysical studies exploring how membrane properties such as fluidity, stiffness and phase separation, affect the propulsion mechanism should therefore prove interesting.

The propulsion mechanism described here appears specific to flagellated bacteria coupled to membrane tubes. It clearly cannot operate for spherical Janus swimmers propelled by phoresis~\cite{howse2007self}. The reason for the absence of vesicle motion with swimming \textit{B.~subtilis} is less clear~\cite{takatori2020active}. A run-and-tumble strain was used in that work, and propulsion in \textit{B.~subtilis} is generated by \textit{multiple} flagella bundles~\cite{Li2011}. Furthermore, the vesicles were produced by electroformation rather than by the inverted emulsion method. Any of these differences might preclude the emergence of vesicle motility (or, indeed, the facile formation of stable membrane tubes). This highlights the specificity of active matter: there is no generic behaviour of swimmers encapsulated in vesicles. Even two bacterial species swimming using flagella bundles produce different effects.

Simple biohybrid systems made of living bacteria encapsulated in synthetic vesicles have been explored for their biotechnological potential in shielding and delivering probiotic bacteria and for biosensing~\cite{cao2019biointerfacial, Elani18,morita2018direct,juskova2019basicles}. Our work adds the possibility of these delivery vehicles becoming self propelled, which might enable more targeted delivery. This could be achieved by exploiting bacterial chemotaxis or by using magnetotactic bacteria guided by magnetic fields~\cite{martel2006controlled}.

Finally, we note that in the field of synthetic life, there has been progress in mimicking biological motility by the `bottom up' generation of self propulsion using entirely synthetic components~\cite{Wang2020}. At the same time, novel biohybrid systems have been generated by coupling synthetic lipid membranes to living cells or isolated cell components~\cite{Elani21}. Using the latter approach, it has long been possible to tow lipid vesicles using external actin polymerization-depolymerization~\cite{Upadhyaya2003}, or by externally attaching bacteria to vesicles ~\cite{dogra2016micro}. The present work reports synthetic vesicle propulsion by \textit{internally} generated forces, in our case by the encapsulation of living cells. 

\section*{Materials and Methods}
All data and analysis codes can be found in the following data repository \url{http://doi.org/10.15128/r2rn3011433}

\subsection*{Bacterial cultures}
We used strain AD83, a smooth-swimming strain derivative of \textit{E.~coli} AB1157. This strain was constructed by deleting the \textit{cheY} gene~\cite{schwarz2016escherichia} and further transformation with plasmid pWR21~\cite{buda2016dynamics}, which expresses eGFP constitutively and confers resistance to Ampicillin. Ampicillin was added to all solutions at a final concentration of \SI{100}{\micro\gram\per\milli\litre}.

Bacteria were plated on lysogeny broth (LB, Miller's formulation) agar plates and grown at \SI{37}{\celsius} to form isolated colonies. LB supplemented with \SI{400}{\milli\Molar} sucrose (LB-sucrose), filter-sterilized using a \SI{0.2}{\micro\metre} filter after sucrose addition, was used thereafter. Unless otherwise specified, \si{\milli\Molar} refers to solute concentrations expressed in millimoles of solute per liter of solvent.

Single colonies were grown overnight in \SI{5}{\milli\litre} of LB-sucrose in an orbital shaker incubator (\SI{37}{\celsius}, $200$ r.p.m.). Then \SI{10}{\milli\litre} of fresh LB-sucrose was inoculated with \SI{100}{\micro\litre} of overnight culture and incubated aerobically for \SI{3.5}{\hour} to an optical density $\textrm{OD}\sim 0.7$. Harvested cells were centrifuged ($6500\times g$, \SI{2}{\minute}), redispersed in fresh LB-sucrose and diluted to $\textrm{OD}=0.3$ with fresh LB-sucrose.

\subsection*{Lipids/oil solution}
POPC dissolved in chloroform was purchased from Avanti Polar Lipids. All other chemicals were purchased from Sigma-Aldrich. The dye 1,1'-dioctadecyl-3,3,3',3'-tetramethylindocarbocyanine perchlorate (DiIC$_{18}$) was stored in ethanol ($\SI{0.58}{\milli\Molar}$ stock solution) and kept in the dark at $\SI{4}{\celsius}$. Lipids/oil solutions were prepared as follows. Thin films of POPC and DiIC$_{18}$ were deposited in separate vials by evaporating the solvents of the respective stock solutions (chloroform and ethanol, respectively) under gentle nitrogen (N$_2$) flow, and dried in vacuum for \SI{2}{\hour}. Light mineral oil was added to the POPC vial to a final POPC concentration of \SI{2}{\milli\gram\per\milli\litre}, and the lipids were dispersed by sonicating for \SI{1.5}{\hour}. The resulting POPC/oil solution was transferred in the vial containing DiIC$_{18}$ and sonicated for a further \SI{30}{\minute} in cold water to disperse the dye (99.5:0.5 POPC:DiIC$_{18}$ molar ratio). Lipids/oil solutions were stored for up to a week at room temperature in N$_2$ and sonicated for $\sim$\SI{10}{\minute} before experiments.

\subsection*{Bacterial encapsulation in GUVs}

Bacteria were encapsulated in GUVs using the inverted-emulsion method~\cite{pautot2003production}. Briefly, a lipid monolayer was formed at an oil/water interface by layering \SI{200}{\micro\litre} of lipids/oil solution on top of \SI{800}{\micro\litre} of aqueous buffer in a \SI{1.5}{\milli\litre} microcentrifuge tube, and the tube was incubated at room temperature for \SI{1}{\hour}. The aqueous buffer (deionized water supplemented with \SI{840}{\milli\Molar} glucose, filter-sterilized) is the outer solution (OS) surrounding the GUVs at the end of the protocol. Protocol steps were timed to ensure that the bacterial suspension was ready shortly before the end of the incubation period.

In a separate tube, \SI{10}{\micro\litre} of bacterial suspension was emulsified in \SI{200}{\micro\litre} of oil/lipids solution by dragging the tube $\sim 15$ times on a tube rack, allowing the formation of lipid-coated water-in-oil droplets. The emulsion was quickly layered on top of the column prepared as described above, and the tube was centrifuged at $1500\times g$ for \SI{5}{\minute} to transfer the droplets through the pre-formed lipid monolayer, thus forming GUVs. Droplet transfer was aided by a small density difference between the OS and the LB-sucrose inner solution (IS) ($\rho_{is}=\SI{1.0593}{\gram\per\centi\metre\cubed}$, $\rho_{os}=\SI{1.0512}{\gram\per\centi\metre\cubed}$). After carefully removing the oil from the tube, GUVs were collected using a micropipette and redispersed in another tube containing fresh OS. That tube was centrifuged at $1000\times g$ for \SI{1}{\minute} and the supernatant discarded to remove most oil droplets and lipid aggregates from the suspension. The final suspension, used in experiments, was obtained by redispersing the washed GUVs into fresh OS.

\subsection*{Imaging and analysis}

Sample chambers ($\sim1\times1\times\SI{0.1}{\centi\metre\cubed}$) were prepared by bonding polydimethylsiloxane (PDMS) spacers to glass coverslips using a plasma cleaner (Zepto, Diener electronic). Just prior to use, chambers were treated for \SI{10}{\minute} with a freshly prepared solution of bovine serum albumin (BSA) dissolved in OS at $10$ \% w/v to minimize GUV adhesion to the glass, and washed thoroughly three times with OS.

Open chambers were filled with \SI{190}{\micro\litre} of GUV suspension. Using a volume of liquid greater than the chambers' volume helped avoid trapping air bubbles while sealing the chambers after evaporation. We used an evaporation-based deflation protocol to obtain softer GUVs. A sample of initial mass $m_i$ with \SI{840}{\milli\Molar} glucose (molar mass $M_g$) in deionized water OS is evaporated to final mass $m_f$:
\begin{equation}
  m_f=\frac{m_i}{\alpha}\frac{1+\alpha c_0 M_{g}}{1+c_0 M_{g}},
\end{equation}
where $c_0=\SI{0.84}{\mole\per\kilo\gram}$ is the initial solute osmolality.

GUVs were imaged near the BSA-treated bottom cover slips using a Nikon TE2000 inverted microscope with Nikon phase contrast objectives (Plan Fluor $20$\texttimes$/0.5$ Ph1 and Plan Apo $60$\texttimes$\lambda/1.4$ Ph3 Oil). Movies were recorded using a scientific Complementary Metal–Oxide–Semiconductor camera (Orca Flash 4.0, Hamamatsu) and Micro-manager~\cite{edelstein2010computer} typically for $3$-\SI{5}{\second}  at $10$ frames per second (fps). High-magnification movies were recorded at $400$~fps for $2$-\SI{5}{\second} to study the motion of helical tubes. We imaged at $1$ fps for \SI{250}{\second} to study GUV motion, tracking them in MATLAB R2020a using a code adapted from reference~\cite{blair2008matlab} and only analysing vesicles showing $> \SI{30}{\second}$ tracks. Instantaneous GUV speeds come from linear fitting of the trajectories over $4$ consecutive frames ($\Delta t= \SI{3}{\second}$ between frames). Whenever possible ($\sim 80\%$ of tracked vesicles), we corrected for advection due to occasional weak residual flows after sealing the sample chambers by tracking a bacteria-free vesicle in the vicinity of motile GUVs and subtracted its drift velocity. The rest of the image analysis was performed using Fiji~\cite{schindelin2012fiji}.

\subsection*{Differential dynamic microscopy}

Bacteria grown in LB-sucrose were centrifuged ($6500\times g$, \SI{2}{\minute}) and diluted to $\textrm{OD}=0.1$ with fresh LB-sucrose, and used immediately to fill rectangle glass capillaries (VitroCom, inner diameters \SI{0.40}{\milli\metre} $\times$ \SI{8}{\milli\metre}) subsequently sealed with petroleum jelly (Vaseline) to prevent evaporation. Bacteria were imaged with a phase contrast objective (Plan Fluor $10$\texttimes$/0.3$) $\approx \SI{150}{\micro\metre}$ above the bottom of the capillaries. \SI{40}{\second} movies were recorded at $100$ fps for  analysis~\cite{wilson2011differential}.  

\section*{Author contributions}

L.L.N., W.C.K.P., and M.S. designed research; A.D. contributed the bacterial strain; L.L.N. and V.A.M. performed experiments; L.L.N. and A.T.B. performed the modeling; L.L.N., A.T.B., W.C.K.P., and M.S. analyzed the data and interpreted the results; L.L.N., A.T.B., A.D., V.A.M., W.C.K.P., and M.S. discussed the results and commented on the manuscript; L.L.N. wrote the first draft; and L.L.N., A.T.B., W.C.K.P., and M.S. wrote the paper.

\section*{Acknowledgments}
We thank Jochen Arlt, Alexander Morozov, Natasha Rigby, Andrew Schofield and Tracy Scott for helpful discussions. This work is a part of a Royal Society Academies Partnership in Supporting Excellence in Cross-disciplinary research award (APX/R1/191020 to M.S., W.C.K.P., and A.T.B.). L.L.N. is funded by Engineering and Physical Sciences Research Council (EPSRC) Centre for Doctoral Training in Soft Matter and Functional Interfaces (Grant EP/L015536/1). A.T.B. is funded by an EPSRC Innovation Fellowship (Grant EP/S001255/1). A.T.B. and W.C.K.P. are funded by the Biotechnology and Biological Sciences Research Council National Biofilms Innovation Centre (Grant BB/R012415/1).

\newpage
\onecolumn

  \begin{center}
{\huge
\textbf{Supporting Information}
}
\end{center}

\renewcommand\thefigure{S\arabic{figure}}
\setcounter{figure}{0} 

\renewcommand\theequation{S\arabic{equation}}
\setcounter{equation}{0}  
\section*{Tubulation analysis}

\subsection*{Statistics}
Evaporation-induced deflation of GUVs leads to a variety of GUV morphologies, with a large variability in the number and length of tubular protrusions observed as well as in the number of bacteria present in these protrusions. The fraction of GUVs $\chi$ that display at least one cell in a tube after deflating the GUVs by a factor $\alpha$ quantifies how easily encapsulated bacteria can extrude a membrane tube and/or enter a spontaneously formed tube. To ensure that any observed difference was not simply linked to aging of the sample, we imaged the $\alpha=1.1$ chamber before the $\alpha=1.05$ one. The results of this analysis are presented in Table~\ref{tubecounting}, which also includes additional information about the radii of the GUVs. GUVs that did not contain bacteria were not included. As expected for GUVs produced by the inverted emulsion method, a wide distribution of radii is obtained. To compare similar GUVs only, we repeated the analysis on a subpopulation of GUVs with radii varying between \SI{5}{\micro\metre} and \SI{10}{\micro\metre}, which represents $\sim45\%$ of the GUV population in each chamber. This analysis yielded almost identical results: $\chi=13\%$ ($\alpha=1$), $\chi=48\%$ ($\alpha=1.05$) and $\chi=68\%$ ($\alpha=1.1$), confirming that deflating the GUVs promotes membrane tubes.

\begin{table}[h]\centering
 \caption{Analysis of GUV populations at different osmotic deflation factors $\alpha$.}
  \label{tubecounting}
  \begin{tabular*}{1\textwidth}{@{\extracolsep{\fill}}cccc}
    \hline
     \multicolumn{1}{c}{${\alpha}$} &
      \multicolumn{1}{c}{${1}$} &
      \multicolumn{1}{c}{${1.05}$} &
      \multicolumn{1}{c}{${1.1}$}\\
      \hline
    ${n}$ & \multirow{1}{*}{$93$} &\multirow{1}{*}{$134$} & \multirow{1}{*}{$114$} \\
    \hline
    ${\chi}$ (\%) & $17$ & $49$ & $67$\\
    \hline
${R_{\rm av}\pm}$ SD & \multirow{1}{*}{$10.8\pm4.5$} &\multirow{1}{*}{$10.3\pm5.4$} & \multirow{1}{*}{$9.6\pm3.8$} \\
     ${R_{\rm med}}$ & \multirow{1}{*}{$10$} &\multirow{1}{*}{$9.1$} & \multirow{1}{*}{$9.5$} \\
     ${[R_{\rm min}~;~R_{\rm max}]}$ & \multirow{1}{*}{$[4.3~;~28.3]$} &\multirow{1}{*}{$[2.9~;~34.5]$} & \multirow{1}{*}{$[3.4~;~23.9]$}\\
    (\si{\micro\metre})& & & \\
     
   \hline
  \end{tabular*}\\
  \begin{tablenotes}
  \item $n$ is the total number of GUVs analysed for each $\alpha$. $\chi$ is the fraction of GUVs displaying one or several bacteria-containing tube(s). $R_{\rm av}$, $R_{\rm med}$, $R_{\rm min}$ and $R_{\rm max}$ are respectively the number averaged, median, minimum and maximum radii recorded for each population. SD: standard deviation.
  \end{tablenotes}
\end{table}

\begin{figure}[h]
\centering
\includegraphics[width=0.65\textwidth]{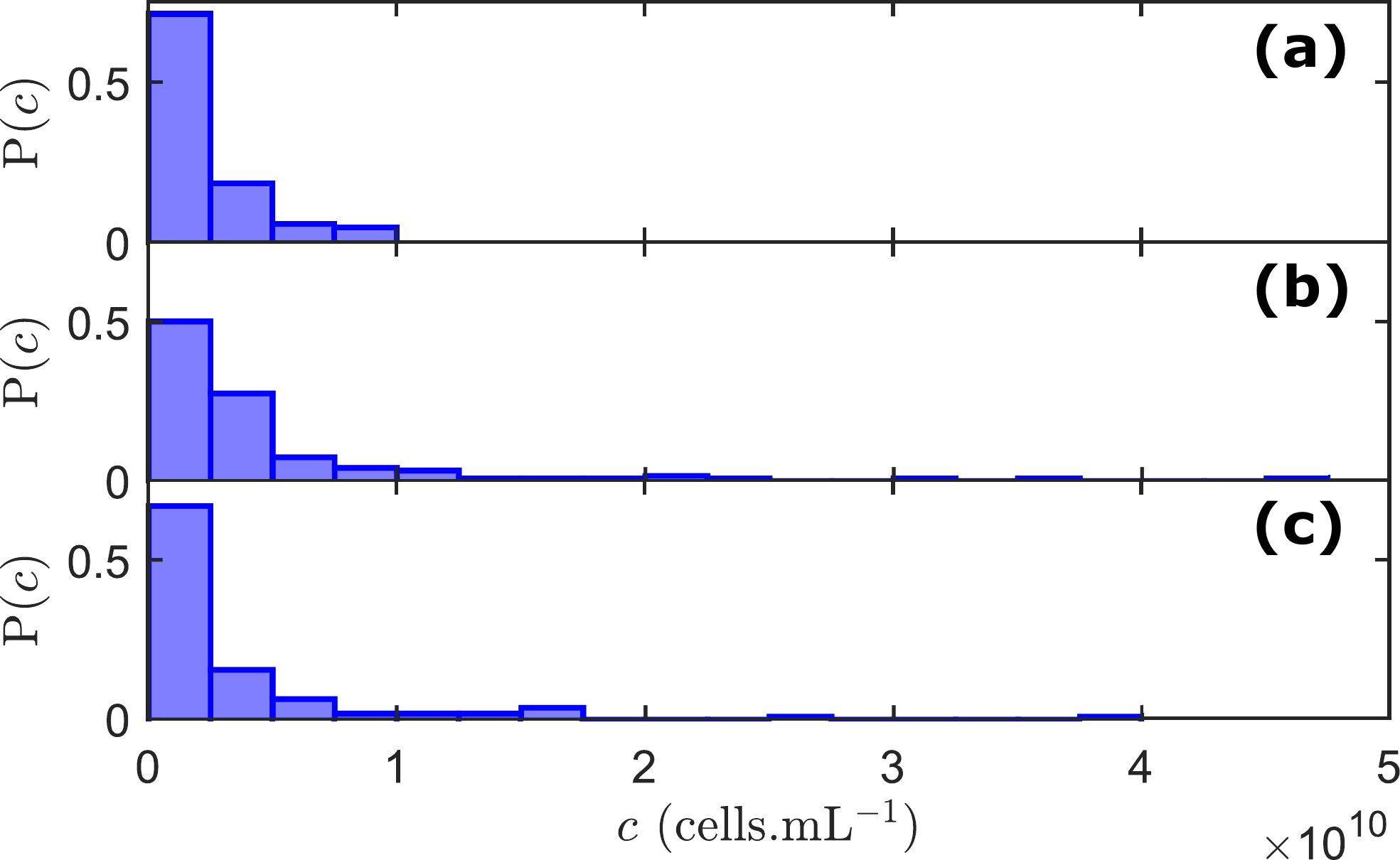}
\caption{\label{Fig_S1}Probability distributions of cell concentrations measured on the GUV populations studied in Table~\ref{tubecounting}. \textbf{(a)} $\alpha=1$, median concentration $c_{\rm m}=\num{1.4e9}$~cells~\si{\per\milli\litre}. \textbf{(b)} $\alpha=1.05$, median concentration $c_{\rm m}=\num{2.6e9}$~cells~\si{\per\milli\litre}. \textbf{(c)} $\alpha=1.1$, median concentration $c_{\rm m}=\num{1.4e9}$~cells~\si{\per\milli\litre}.}
\end{figure}

For each GUV included in the analysis of Table~\ref{tubecounting}, we also estimated the concentration $c$ of encapsulated bacteria by counting the cells present in the vesicle. Despite the relatively large uncertainty associated with this counting method -- especially for high bacterial concentrations -- we observed that $\chi$ did not seem to depend on $c$ for the range of concentrations used in our experiments. The number averaged concentrations (sum of all measured concentrations divided by the number of analysed GUVs) were $c=(2.2\pm2.4)\times10^9$~cells~\si{\per\milli\litre}, $c=(5.6\pm8.7)\times10^9$~cells~\si{\per\milli\litre} and $c=(3.7\pm5.7)\times10^9$~cells~\si{\per\milli\litre}  (mean $\pm$ standard deviation) for $\alpha=1$, $\alpha=1.05$ and $\alpha=1.1$ respectively. The cell concentration distributions are not symmetric, Fig.~\ref{Fig_S1}, so that the number averaged concentrations tend to be shifted towards higher values due to the presence of relatively rare GUVs that have a very high concentration of bacteria. We thus also report median concentrations: $c_{\rm m}=\num{1.4e9}$~cells~\si{\per\milli\litre}, $c_{\rm m}=\num{2.6e9}$~cells~\si{\per\milli\litre} and $c_{\rm m}=\num{1.4e9}$~cells~\si{\per\milli\litre} for $\alpha=1$, $\alpha=1.05$ and $\alpha=1.1$ respectively. For a GUV with $R=\SI{10}{\micro\metre}$, these values correspond to a median of $6$ to $11$ encapsulated bacteria.

\subsection*{Tube formation and GUV propulsion require motile encapsulated cells}
A recent study using a pathogenic strain of \textit{E.~coli} showed that bacteria can adhere to the lipid membrane of GUVs~\cite{cazzola2020impact}. We exposed empty GUVs to an outer medium with motile bacteria (OD~$=0.3$), and saw no structures resembling bacteria-containing tubes. Also, no GUV motion was observed for bacteria-free GUVs exposed to non-encapsulated bacteria.

We also sought to confirm that only \textit{motile} bacteria could extrude membrane tubes or enter spontaneously formed membrane protrusions. We encapsulated dead \textit{E.~coli} cells, obtained by heating a culture at \SI{60}{\celsius} for \SI{15}{\minute} followed by the  evaporation protocol. No bacteria were observed in membrane tubes in these conditions. Even if empty membrane tubes sometimes form spontaneously during the evaporation protocol, active swimming is required for bacteria to be observed in membrane tubes.

\section*{Absence of correlation between $f$ and $R$}

Fig.~\ref{Fig_Force} shows that the propulsive force exerted by a single cell $f$ (deduced from measurements as detailed in the main text) is uncorrelated with the GUV radius $R$ (Pearson correlation coefficient $r=0.13$).

\section*{Differential dynamic microscopy analysis of freely-swimming bacteria in LB-sucrose}

Differential dynamic microscopy (DDM) can characterise the speed distribution of motile \textit{E. coli} by assuming that it  takes Schulz form~\cite{martinez2012differential}: 
\begin{equation}
\label{EqS1}
  P(v)=\frac{v^Z}{Z!}\left(\frac{Z+1}{\bar{v}}\right)^{Z+1}e^{-\frac{v}{\bar{v}}(Z+1)},
\end{equation}
where $\bar{v}$ is the mean of the speed distribution and $Z$ is related to the variance $\sigma_v^2$ of $P(v)$ via $\sigma_v^2=\frac{\bar{v}^2}{Z+1}$. We used DDM to characterize an aerobic suspension of freely-swimming \textit{E. coli} in LB-sucrose and measured $\bar{v}=15.9\pm\SI{1.0}{\micro\metre\per\second}$ and $\sigma_v=6.2\pm\SI{0.9}{\micro\metre\per\second}$, Fig.~\ref{Fig_distrib}(a). Uncertainties come from the fits of the DDM analysis. We then estimate the thrust force distribution for freely-swimming cells by multiplying the speeds of Fig.~\ref{Fig_distrib}(a) by $\frac{k_BT}{D}$, where $D$ is the isotropic diffusion coefficient of an \textit{E. coli} cell. For non-motile \textit{E. coli} in motility buffer, $D_0\sim\SI{0.3}{\micro\metre\squared\per\second}$~\cite{wilson2011differential}. We use $D=\frac{2}{3}D_0$ to account for the increased viscosity from sucrose addition. The force distribution of freely-swimming cells is shifted towards higher values compared to the one obtained from bacteria in membrane tubes, as expected, see main text and Fig.~\ref{Fig_distrib}(b). However, both distributions have similar shapes and coefficients of variation ($c_v=0.51$ for bacteria in membrane tubes compared to $c_v=0.39$ for freely-swimming cells), confirming that biological variability contributes significantly to the spread in $f$ observed in Fig.~4(a) of the main text.

\section*{Drag on empty portions of tube}
Our model presented in the main text takes into account the drag on the vesicle, $\xi_v$, and that of each bacterium in the tube, $\xi_b$, but it neglects the drag contribution from the empty portion of the tube, $\xi_t$, for the reasons stated below.

The drag coefficient of a cylindrical tube of length $L_t$ and diameter $D_t$ moving along its long axis is $\xi_t=\frac{2\pi\eta L_t}{\ln{\left(2L_t/D_t\right)}-0.807}$~\cite{cox1970motion}.  Averaging over all tracked vesicles with experimentally measured tube lengths (subtracting a length of $\SI{3}{\micro\metre}$ per cell body) and vesicle radii, we find that the tube drag is small compared to that of the spherical vesicle, i.e. $0.19<\frac{\xi_t}{\xi_v}=\frac{\xi_t}{6\pi\eta R}<0.28$ for $\SI{0.05}{\micro\metre}<D_t<\SI{0.4}{\micro\metre}$. (Note that the actual tube radii are below our optical resolution. Therefore we are exploring a range of plausible values). More importantly, we find no dependency of $\xi_t$ on the number of bacteria present in the tube (the average $\frac{\xi_t}{\xi_v}$ for $D_t=\SI{0.4}{\micro\metre}$ varies between 0.25 and 0.31 for all $\mathcal{N}$).

For comparison, the drag coefficient of a single bacterial cell is also relatively small; averaged over all GUVs it is $\frac{\xi_b}{\xi_v}=0.10$. However, we must include the drag on the bacteria to avoid creating an $\mathcal{N}$-dependent error when plotting the force as a function of $\mathcal{N}$ in Fig.~4. This is not the case for $\xi_t$, the omission of which results in a slight underestimation of the calculated force but does not affect the linear trend between the propulsive force and $\mathcal{N}$. This is demonstrated in Fig.~\ref{Force_tube_drag}, which compares the forces calculated with and without taking $\xi_t$ into account for $D_t=\SI{0.4}{\micro\metre}$. The fitted force $\langle f \rangle = \SI{0.13}{\pico\newton}$ (regression coefficient~= 0.7) is similar to the value obtained in Fig.~4(b) without tube drag.

\section*{Variations in helix thickness weakly affect the propulsive force}

Tube-to-tube variations in the thickness $2a$ of the helical portions of tube may affect the magnitude of the propulsive force and may contribute to the variability in $f$. Here, we show that this contribution is small over the plausible range of $a$ in our experiments, $\SI{0.05}{\micro\metre}<2a<\SI{0.4}{\micro\metre}$.

We estimate the effect of a varying thickness on the generated thrust force using resistive force theory (RFT)~\cite{chattopadhyay2006swimming,rodenborn2013propulsion}. We use Lighthill’s drag coefficients $C_n$ and $C_t$ to estimate the thrust force generated by the rotation of a helix with pitch $p=\SI{2.3}{\micro\metre}$, length $l=\SI{5}{\micro\metre}$, helical diameter $d=\SI{0.4}{\micro\metre}$ and rotation speed $\omega=2\pi f_{\Omega}$ with $f_{\Omega}=\SI{100}{\hertz}$:
\begin{equation}
\label{EqS2}
  F_{\rm RFT}=\frac{\sin\phi}{2}\omega ld\left(C_n-C_t\right),
\end{equation}
with $C_n=\frac{4\pi\eta}{\ln{\frac{cp}{a\cos\phi}}+1/2}$, $C_t=\frac{2\pi\eta}{\ln{\frac{cp}{a\cos\phi}}}$, $c=0.18$ and pitch angle $\phi$ given by $\tan\phi=\pi d/p$. As shown in Fig.~\ref{Fprop_vs_Rt}, $F_{\rm RFT}$ varies weakly with $a$ over the considered range. Note that RFT has been shown to significantly overestimate the thrust force~\cite{rodenborn2013propulsion}, which explains why the force predicted here is larger than the thrust force of free-swimming \textit{E. coli}. This still allows us to probe the dependency of the force on $a$ and to show that we expect it to be negligible compared to other sources of variability, such as the biological variability demonstrated by DDM.

\section*{Tube formation in different experimental conditions}
In our single-lipid GUVs, spontaneous curvature could arise from salt/sugar compositional asymmetry across the membrane~\cite{bhatia2020simple,karimi2018asymmetric}. To assess the reproducibility of the formation of active membrane tubes, we tested whether tube formation depends on the presence of specific molecules or on asymmetric conditions potentially leading to spontaneous curvature. We first verified that tube formation is unaffected in the absence of dye in the membrane, then tested the experimental conditions detailed below.

\subsection*{Symmetric, growth medium based conditions}
We tested whether asymmetric sugar and salt concentrations are required for the formation of thin tubes. We encapsulated motile bacteria in GUVs using the protocol detailed in the main text, and diluted the GUV suspension 50 times in LB supplemented with \SI{400}{\milli\Molar} sucrose. After dilution, the outer medium thus consisted of slightly diluted LB supplemented with \SI{392}{\milli\Molar} sucrose and \SI{17}{\milli\Molar} glucose, very similar to the inner solution. The previously described evaporation protocol gave (sealed) samples at $\alpha=1.1$. These conditions gave tubes similar to those observed in the main experiments, with many tubes being visibly thinner than the theoretical estimate of $R_{t,min}\sim\SI{1.3}{\micro\metre}$  given in the main text, (Fig.~\ref{Fig_controlexp}(a)-(c)). This indicates that salt and sugar concentration asymmetry across the membrane is unlikely to be the main driver for tube formation. Taking advantage of the low density difference between the GUVs and the outer medium in these conditions, we also confirmed that bacteria-containing tubes can be seen while the GUVs are far from the BSA-coated substrate.

\subsection*{Defined buffers with sugar asymmetry} Using growth medium for the inner solution provides nutrients to the bacteria and minimizes the need to wash and transfer them into different buffers before encapsulation. However, LB is an undefined medium containing amino acids and peptides. We thus tested if these potentially membrane-active molecules are required for the formation of thin tubes by repeating the experiments with deionized water supplemented with \SI{212}{\milli\Molar} sucrose and glucose respectively for the inner and outer solutions. Briefly, bacteria were grown in pure LB, harvested after \SI{3.5}{\hour} of growth at \SI{37}{\celsius}, washed twice by centrifugation (6500~g, \SI{2}{\minute}) and diluted in inner solution to OD=0.3. Encapsulation in GUVs followed the protocol detailed in the main text, except that the column was centrifuged at only $400\times g$ because of the larger density difference between the inner and outer media in these conditions. The imaging chamber was sealed at $\alpha=1.1$. We saw tubes similar to those observed in the main experiments, and the same coupling between tubes and flagella bundles leading to GUV motion  (Fig.~\ref{Fig_controlexp}(d)-(h)). We conclude that the peptides contained in LB do not play a prominent role in tube formation.

Note that different sugar solutions across the membrane can generate spontaneous curvature. The sucrose and glucose concentrations used here match those used in~\cite{bhatia2020simple}, which generated a spontaneous curvature $m=\SI{1.3}{\per\micro\metre}$. In the presence of spontaneous curvature, the force needed to pull a membrane tube becomes $f=2\pi\sqrt{2\kappa\sigma}-4\pi\kappa m$ and the equilibrium tube radius is $R_t\sim\sqrt{\frac{\kappa}{2\left(\sigma+2\kappa m^2\right)}}$. Using values from the main text, we find a theoretical minimum radius $R_{t,min}\sim\SI{0.3}{\micro\metre}$, \textit{i.e.}, thinner than the bacteria and in agreement with the pictures displayed in Fig.~\ref{Fig_controlexp}(d)-(h). However, further control experiments detailed in the next paragraph show that a sugar-induced positive spontaneous curvature is not needed to obtain thin tubes. 

\subsection*{Defined buffers, symmetric conditions} As a final control, we repeated the experiment with defined, symmetric conditions. GUVs produced with the protocol described in the previous paragraph were diluted 10 times in deionized water supplemented with \SI{212}{\milli\Molar} sucrose. The outer solution was thus almost identical to the inner medium (deionized water supplemented with \SI{212}{\milli\Molar} sucrose). Once again, bacteria were observed in thin membrane tubes (Fig.~\ref{Fig_controlexp}(i)\&(j)), indicating that spontaneous curvature is not needed to explain the formation of these tubes.

In sum, these control experiments indicate that a spontaneous curvature of molecular origin (peptides and/or compositional asymmetry across the membrane) is unlikely to be the main factor explaining the spontaneous formation of thin empty tubes or the thinner than expected diameter of bacteria-containing tubes. We cannot fully exclude a potential effect of leftover oil or defects in the membrane but, as explained in the main text, an initial area difference between the inner and outer leaflets of the membrane could be the main factor aiding tube formation in our system.

\newpage

\begin{figure}[h]
\centering
\includegraphics[width=0.75\textwidth]{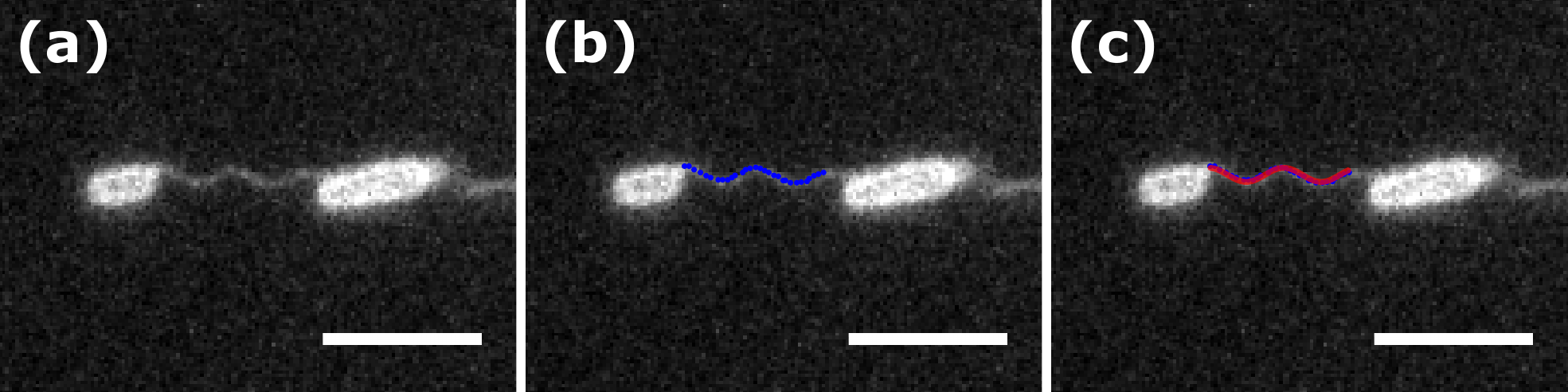}
\caption{Image analysis of the flagella bundle. \textbf{(a)} Raw image showing a helical, horizontally oriented portion of tube at the back of a small cell, immediately followed by another larger cell. \textbf{(b)} Manually extracted tube profile (blue points). \textbf{(c)} Fit of the tube profile with a sinusoidal function (superimposed red curve). The function used is $y=\frac{d}{2}\times sin\left(\frac{2\pi x}{p}+c_1\right)+c_2$ and directly returns the diameter $d=\SI{0.44}{\micro\metre}$ and pitch $p=\SI{2.36}{\micro\metre}$ of the helical bundle. (Scale bars, \SI{5}{\micro\metre}.)}
\end{figure}

\begin{figure}[h!]
\centering
\includegraphics[width=0.5\textwidth]{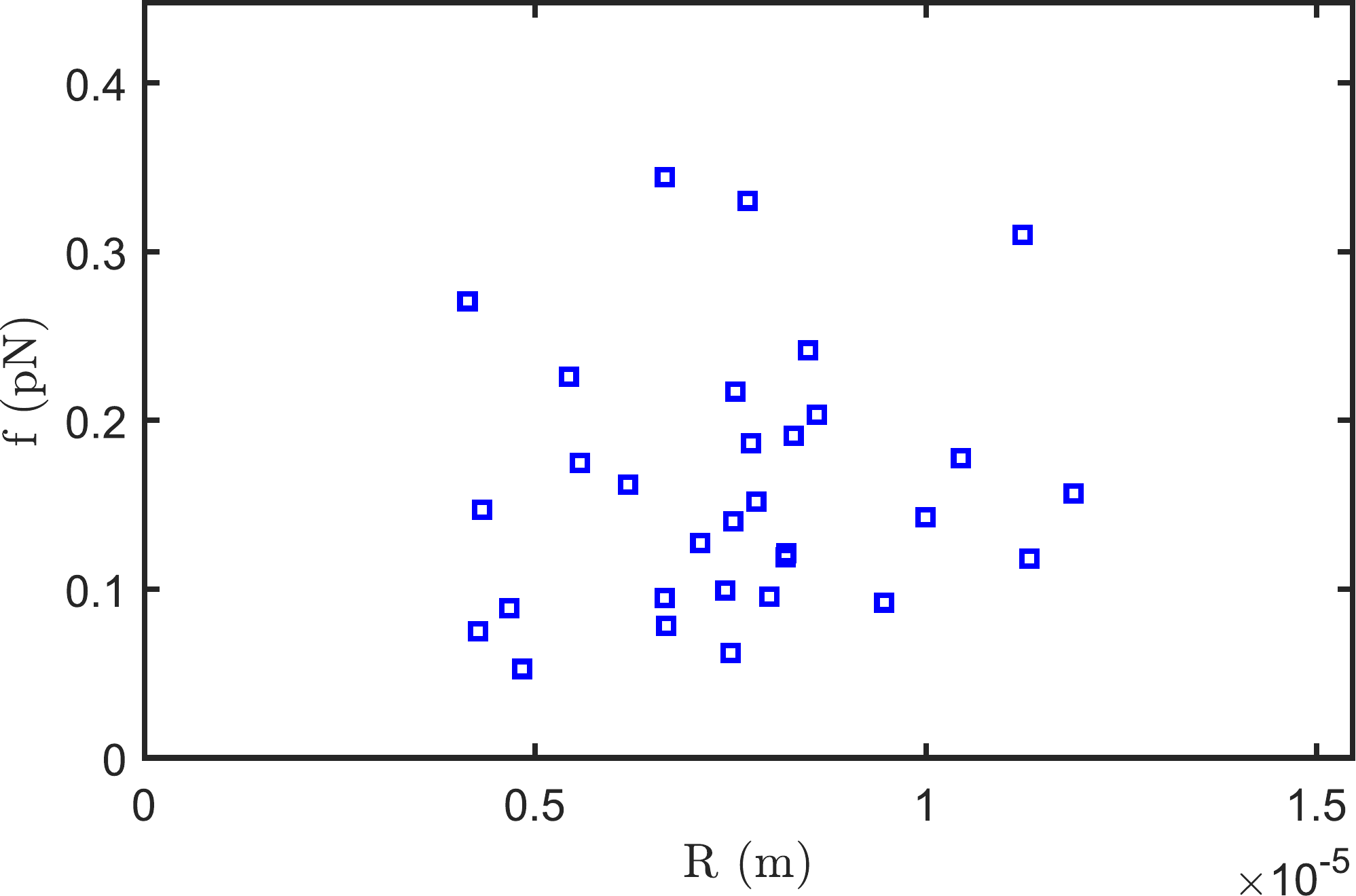}
\caption{\label{Fig_Force} Calculated propulsive force plotted as a function of GUV radius for all GUVs propelled by a single bacterium. As expected, no correlation is observed between the two variables.}
\end{figure}

\pagebreak

\begin{figure}[h!]
\centering
\includegraphics[width=0.8\textwidth]{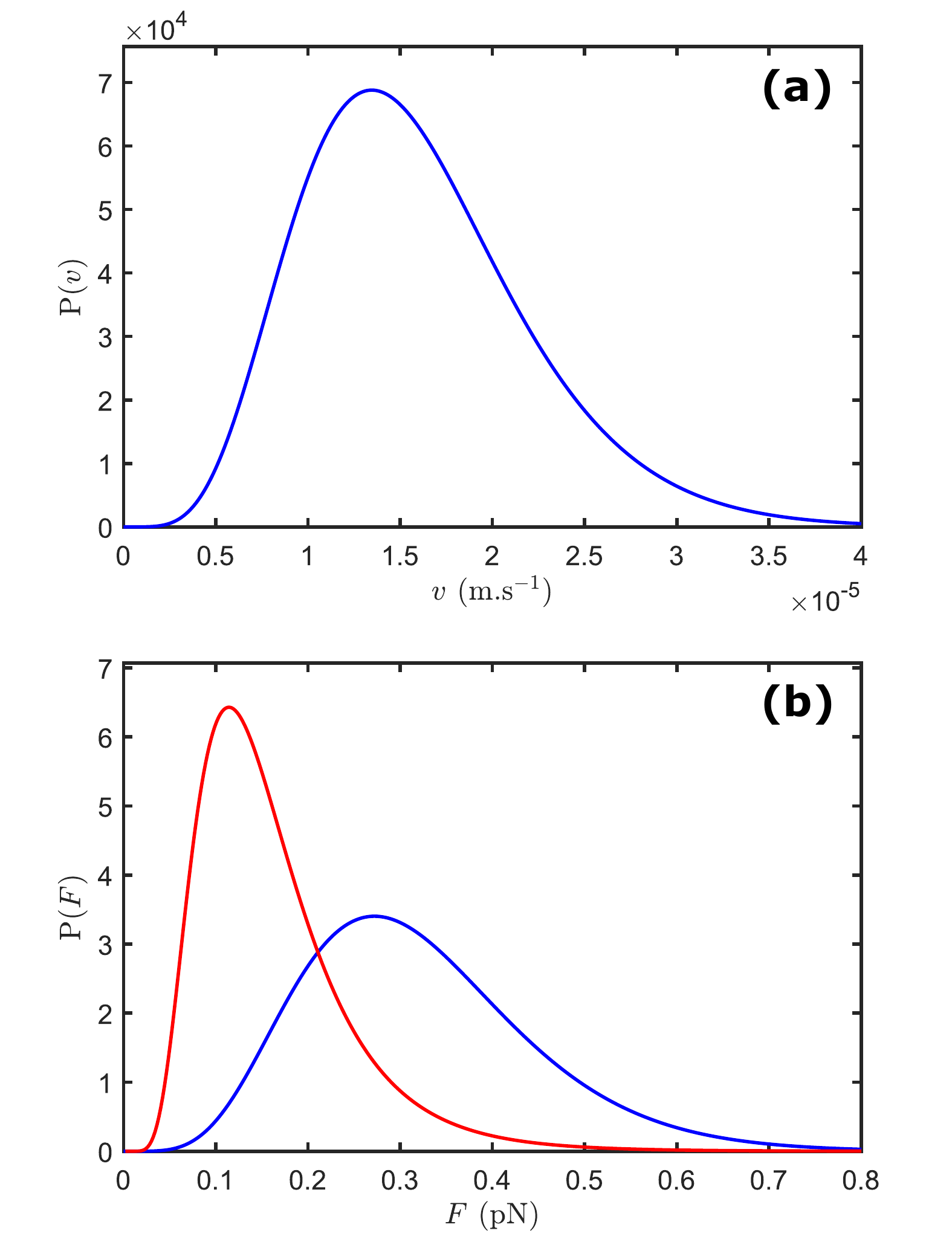}
\caption{\label{Fig_distrib} DDM analysis of freely-swimming cells. \textbf{(a)} Schulz distribution of swimming speeds obtained for a suspension of freely-swimming \textit{E. coli} in LB-sucrose, with $\bar{v}=\SI{15.9}{\micro\metre\per\second}$ and $\sigma_v=\SI{6.2}{\micro\metre\per\second}$. \textbf{(b)} Comparison between the thrust force distribution obtained for freely-swimming bacteria (blue line, derived from the speed distribution with $\bar{F}=\SI{0.32}{\pico\newton}$ and $\sigma_F=\SI{0.13}{\pico\newton}$) and for bacteria in membrane tubes (red line, corresponding to the log-normal distribution of Fig.~4(a) in the main text). Both distributions have similar coefficients of variation: $c_v=0.51$ for bacteria in membrane tubes compared to $c_v=0.39$ for freely-swimming cells.}
\end{figure}

\pagebreak

\begin{figure}[h!]
\centering
\includegraphics[width=\textwidth]{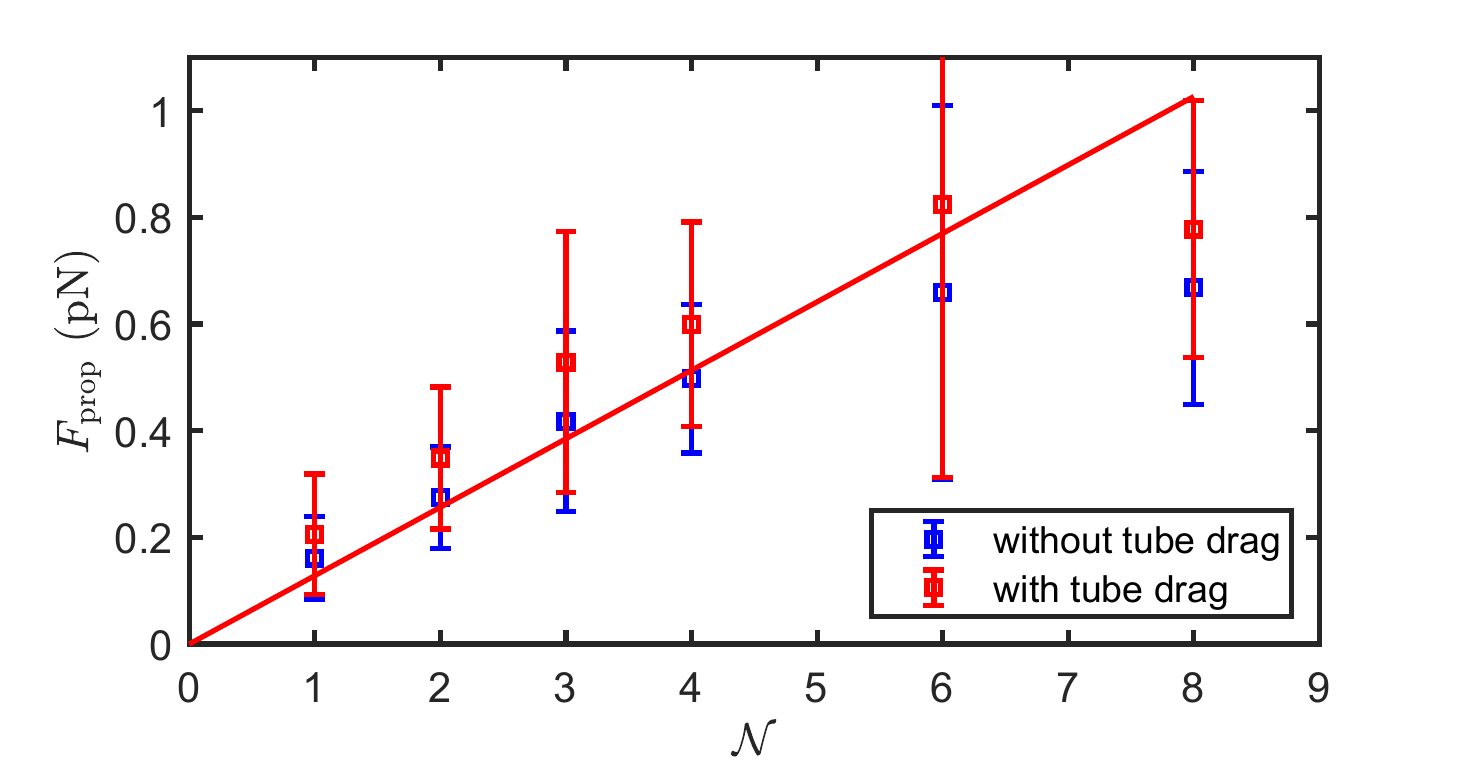}
\caption{\label{Force_tube_drag} Average value of the total propulsive force $F_{\rm prop}$ generated by bacteria as a function of the number of bacteria $\mathcal N$ in the tubes, with and without taking into account tube drag. Red points are calculated using an upper bound for the drag on the tube, $D_t=\SI{0.4}{\micro\metre}$. The red line is a linear fit of the red points weighted by the inverse of the variance of the data, with a fitted slope $\langle f \rangle =\num{0.13}\mathcal{N}~\si{\pico\newton}$ ($R^2=0.7$). Error bars correspond to standard deviations.}
\end{figure}

\begin{figure}[h!]
\centering
\includegraphics[width=\textwidth]{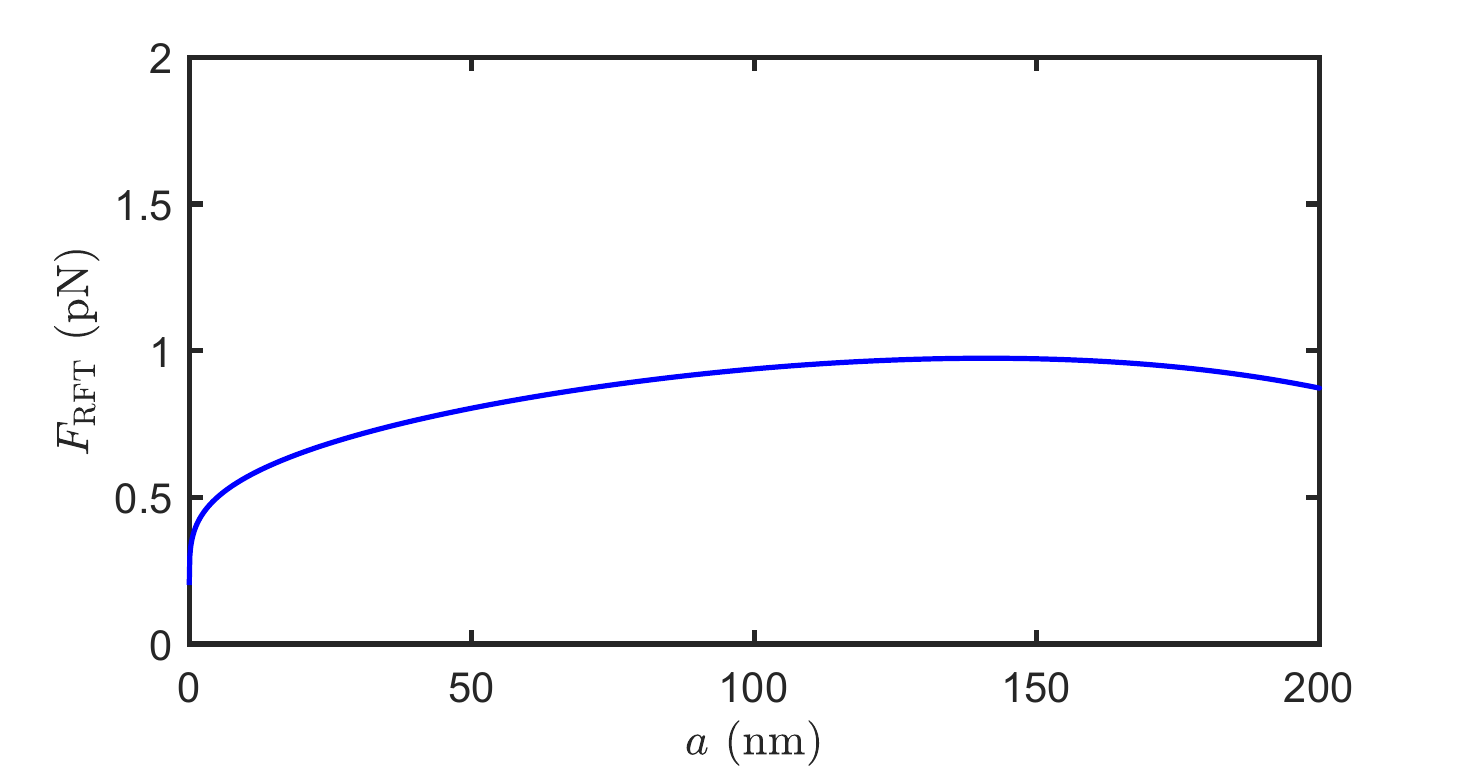}
\caption{\label{Fprop_vs_Rt}The thrust force $F_{\rm RFT}$ calculated for a helix of thickness $2a$ using resistive force theory varies weakly with $a$ over the plausible range covered in experiments, $\SI{25}{\nano\metre}<a<\SI{200}{\nano\metre}$.}
\end{figure}

\pagebreak
\begin{figure}[h!]
\centering
\includegraphics[width=\textwidth]{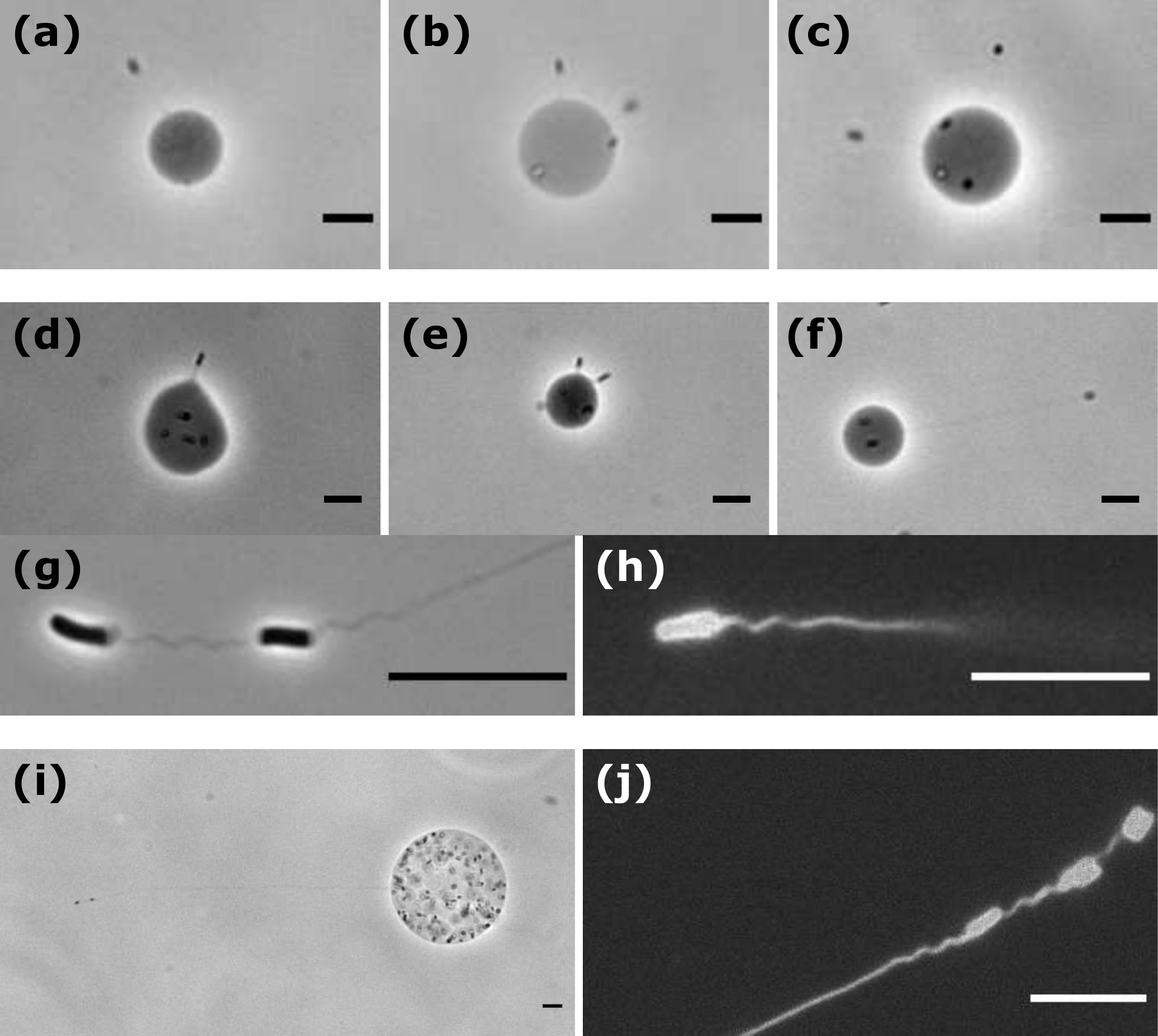}
\caption{\label{Fig_controlexp} Thin tubes coupling with flagella obtained in three independent experiments using different IS and OS (see text for details). \textbf{(a)-(c)} Symmetric conditions using sucrose-supplemented LB. \textbf{(d)-(h)} Asymmetric conditions using an IS made of sucrose supplemented deionized water and an OS made of glucose supplemented deionized water. \textbf{(i),(j)} Symmetric conditions using sucrose-supplemented deionized water. (Scale bars, \SI{10}{\micro\metre}.)}
\end{figure}

\pagebreak

\bibliographystyle{pnas-new}
\bibliography{main}
\end{document}